\documentclass[preprint,authoryear,12pt]{elsarticle}

\usepackage{array}
\usepackage{multirow}

\usepackage[ps2pdf,%
a4paper=true,%
breaklinks=true,%
colorlinks=true,%
pdfauthor={Ferrand & Safi-Harb},%
pdftitle={A Census of High-Energy Observations of Galactic Supernova Remnants}%
]{hyperref}

\journal{Advances in Space Research}

\begin{document}


\begin{frontmatter}

\title{A Census of High-Energy Observations\\ of Galactic Supernova Remnants}

\author{Gilles Ferrand\corref{cor}\fnref{footnoteGilles}}
\fntext[footnoteGilles]{CITA National Fellow}
\ead{gferrand@physics.umanitoba.ca}
\cortext[cor]{Corresponding author}

\author{Samar Safi-Harb\fnref{footnoteSamar}}
\fntext[footnoteSamar]{Canada Research Chair}
\ead{samar@physics.umanitoba.ca}

\address{Department of Physics and Astronomy, University of Manitoba,\\ Winnipeg, MB, R3T~2N2, Canada\\
\url{http://www.physics.umanitoba.ca/snr}}

\begin{abstract}

We present the first public database of high-energy observations of all known Galactic supernova remnants (SNRs). 
In section~\ref{sec:Introduction} we introduce the rationale for this work motivated primarily by studying particle acceleration in SNRs, 
and which aims at bridging the already existing census of Galactic SNRs (primarily made at radio wavelengths)
with the ever-growing but diverse observations of these objects at high-energies (in the X-ray and $\gamma$-ray regimes). 
In section~\ref{sec:Access} we show how users can browse the database using a~dedicated web front-end (\url{http://www.physics.umanitoba.ca/snr/SNRcat}). 
In section~\ref{sec:Statistics} we give some basic statistics about the records we have collected so far, 
which provides a summary of our current view of Galactic SNRs. 
Finally, in section~\ref{sec:Perspectives}, we discuss some possible extensions of this work. 
We believe that this catalogue will be useful to both observers and theorists, 
and timely with the synergy in radio/high-energy SNR studies as well as the upcoming new high-energy missions. 
A~feedback form provided on the website will allow users to provide comments or input, 
thus helping us keep the database up-to-date with the latest observations.

\end{abstract}


\end{frontmatter}

\parindent=0.5 cm


\section{Introduction\label{sec:Introduction}}

The initial motivation for the census of high-energy observations of Supernova Remnants (SNRs) is the study of particle acceleration:
SNRs are believed to be the main production sites of high-energy cosmic rays in the Galaxy, 
and much progress has been achieved in the past decade with the advance of high-energy X-ray and gamma-ray observatories 
(such as Chandra, XMM-Newton, Suzaku, Fermi, H.E.S.S.) and the observation of hard non-thermal radiation from many SNRs. 
Observations at high-energies have been particularly important to assess the acceleration of protons, 
which make the most important part of cosmic radiation, 
but unfortunately radiate far less efficiently than electrons. 
Observations of thermal X-ray emission are also very useful to understand how particles get accelerated by shocks, and modify them.
The purpose of this work is not to discuss all these issues in detail, 
nor to review them in depth or provide a modelling of a specific remnant, 
but rather to offer a global view of the current status of high-energy observations of SNRs. 
This will hopefully help a broad community, 
both theoreticians to apply their models or design new numerical simulations, 
and observers to plan future observations or design new instruments. 
The wealth of existing state-of-the-art X-ray and gamma-ray missions 
and the planning of future high-energy missions 
(NuSTAR, ASTROSAT, eROSITA, ASTRO-H, GEMS and HXMT in X-rays expected to be launched by or during 2014, 
and the CTA in TeV gamma-rays currently in the preparatory phase) 
make this work timely and highly desired by the SNR, Interstellar Medium (ISM), and cosmic ray communities.

\subsection{Existing resources\label{sub:Existing-resources}}

This work builds upon an ensemble of previous studies (of specific objects or at specific wavelengths), 
which is quite vast but rather diverse, and which we here combine, update, and/or summarize 
to provide a global view of all types of SNRs at high-energies. 
We certainly do not plan to replace them, but to offer a consistent, unified way 
to access them all and keep the information up-to-date 
-- again with the focus on the high-energy X-ray and gamma-ray observations of Galactic SNRs.

\subsubsection{Catalogues of objects}

The reference for Galactic SNRs is Dave A. Green's \emph{Catalogue of Galactic Supernova Remnants}, 
available on the web as a dedicated web site\footnote{\url{http://www.mrao.cam.ac.uk/surveys/snrs/}}, 
and through the VizieR service\footnote{\url{http://vizier.u-strasbg.fr/viz-bin/VizieR?-source=VII/253}}. 
The catalogue aims at listing all known Galactic remnants, as mostly identified from radio observations. 
It was first published in~\cite{Green1984a} and updated a number of times in the past 25 years, 
with the last public revision made in~\cite{Green2009b}.

We also found useful to integrate the data from two population studies:
the \emph{list of Galactic SNRs Interacting with Molecular Clouds}
compiled by Bing Jiang\footnote{\url{http://astronomy.nju.edu.cn/~ygchen/others/bjiang/interSNR6.htm}} 
(first published as an appendix in~\citealt{Jiang2010a}),
and the \emph{census of the youngest Galactic SNRs} by Matthieu Renaud
(presented in a workshop talk available online\footnote{\url{http://www.cenbg.in2p3.fr/heberge/MSPWorkshop/IMG/pdf/M-_Renaud.pdf}}).
Young remnants are probably the most studied and well known objects,
and it is believed that the highest energies of accelerated particles are reached in the early phases of expansion. 
On the other hand, molecular clouds are particularly interesting as they greatly enhance the emission induced by accelerated protons. 

Finally, even though we are mostly concerned here with SNRs understood
as the diffuse emission from the expanding ejecta and shocked ISM
rather than the compact object left behind the explosion, 
we find it important to provide the context in which each SNR evolves 
(especially when it comes to the interpretation of high-energy emission 
and the study of neutron stars-SNR associations), 
so we cross-checked our entries with two catalogues of related objects: 
the \emph{Pulsar Wind Nebula Catalog} 
(not up-to-date but first created online around 2005 by Mallory Roberts\footnote{\url{http://www.physics.mcgill.ca/~pulsar/pwncat.html}})
and the \emph{SGR/AXP Catalog} 
(which is maintained online by the McGill Pulsar group\footnote{\url{http://www.physics.mcgill.ca/~pulsar/magnetar/main.html}}).
We have also cross-checked our entries with the PWNe list of \cite{Kargaltsev2008a}.

\subsubsection{Catalogues of observations}

Observations are usually presented in the context of a given experiment, especially in online materials. 
Some observatories offer a wealth of information, notably the Chandra satellite, 
which provides a catalogue of SNRs\footnote{\url{http://hea-www.cfa.harvard.edu/ChandraSNR/}} 
(in both the Galaxy and the Magellanic Clouds), 
and has a dedicated section in its photo album\footnote{\url{http://chandra.harvard.edu/photo/category/snr.html}}. 
The XMM-Newton satellite also has a section dedicated for SNRs 
in its web gallery\footnote{\url{http://xmm.esac.esa.int/external/xmm_science/gallery/public/level2a.php?p=0&cat=3&subcat=2}}, 
but does not offer a comprehensive list of observed objects. 
Compilations of observations of SNRs are also available for two previous major X-ray satellites, ASCA and ROSAT: 
ASCA provides an atlas\footnote{\url{http://www.astro.columbia.edu/~eric/atlas/asca/snrs/index_by_name.html}} 
and ROSAT an image gallery\footnote{\url{http://www.mpe.mpg.de/xray/wave/rosat/gallery/images/sn_snr.php}}. 
There is also a list of SNRs observed by an early pioneer in this energy domain, 
the EINSTEIN satellite\footnote{\url{http://hea-www.harvard.edu/~slane/snr/cats/einstein_list/tbl1.html}}.

In the $\gamma$-ray domain, the past few years have witnessed a synergy and a bloom of observations 
obtained with a number of space- and ground-based missions. 
The H.E.S.S. Cherenkov telescopes array has a strong online presence, 
with a source catalogue\footnote{\url{http://www.mpi-hd.mpg.de/hfm/HESS/pages/home/sources/}} 
including several SNRs, as well as a collection of {}``source of the month'' articles%
\footnote{\url{http://www.mpi-hd.mpg.de/hfm/HESS/pages/home/som/}}. 
The VERITAS experiment offers a growing list of major results%
\footnote{\url{http://veritas.sao.arizona.edu/veritas-science/veritas-results-mainmenu-72}}, 
including a few SNRs. 
In this high-energy domain, one can also find a few catalogues that summarize all available observations 
-- for all instruments and possible types of sources. 
The most recent and up-to-date one is TEVcat, 
which offers a powerful web interface\footnote{\url{http://tevcat.uchicago.edu/}} 
that was a source of inspiration. 
In the GeV range, we checked the latest catalogues 
from the Fermi~\citep{Abdo2009b,Abdo2010d,Fermi-LAT-Collaboration2011a}
and Agile~\citep{Pittori2009a} missions. 

At the other end of the electromagnetic spectrum, in the radio regime,
many telescopes have produced catalogues, and some dedicated to SNRs 
(such as MOST\footnote{\url{http://www.physics.usyd.edu.au/sifa/Main/MSC}}). 
They were not directly used in this work, 
considering that they are already summarized by D.A.~Green's work; 
our focus also being on the high-energy observations. 
We have however cross-checked our list of SNRs with the one published~\citep{Kothes2006a} 
from the Canadian Galactic Plane Survey\footnote{\url{http://www.ras.ucalgary.ca/CGPS/}}.

Finally we note that there is a multi-wavelength atlas of SNRs: 
the \emph{Atlas of radio, X-ray and optical maps of Galactic SNRs}, 
available online\footnote{\url{http://www.sao.ru/cats/~satr/SNR/snr_map.html}} 
as part of the CATS database. 
It is a database of observation maps which gives direct access to a large collection of images. 
However, it was last updated in 2000, just before the boom of high-energy observations.

\subsection{Aims\label{sub:Aims}}

Our own work was developed with a few key goals in mind:
\begin{itemize}

\item focus on \textbf{high energies}: 
Green's catalogue is already complete for radio observations, 
except for some discoveries since 2009 (Green's list currently includes 274 SNRs). 
It~has been used as the base list for our catalogue. 
D.~A.~Green includes references to observations at all wavelengths, 
but he is mostly concerned with SNR identification, typing, and overall properties. 
While these characteristics are also of interest to us and are highlighted/updated in our catalogue, 
we are concerned with particle acceleration in SNRs up to TeV or even PeV energies 
as revealed or hinted for by the broadband X-ray and gamma-ray emission.

\item provide a \textbf{unified view} of SNRs: 
all observations from the major relevant high-energy observatories are presented together for the first time, 
in a single framework independent of any particular instrument.

\item provide an \textbf{up-to-date} catalogue: 
D.~A.~Green updates his catalogue every few years, 
which is no longer sufficient to keep pace with the surge in X-ray and $\gamma$-ray observations 
(new ones are now published weekly, if not daily)\footnote{After the initial lengthy phase of data collection, 
it will be easier for us to update the catalogue regularly, especially as the community uses it and provides us with feedback.}. 
This implies that editing or adding new data should be a quick process,
and that any additional layers for accessing the data (such as a web interface) 
must be re-generated automatically and instantly.

\item provide a catalogue that is \textbf{easy to manipulate:} 
we use a data structure which allows basic operations such as sorting or filtering.
Data can then be reformatted or plugged into other systems, according to our needs.

\end{itemize}
Our objective is to provide a catalogue that is most useful 
for the planning of future observations with existing and upcoming missions
(across the electromagnetic spectrum), 
and for the designing of models and simulations.


\section{Web access\label{sec:Access}}

All the data are stored inside a MySQL relational database, 
hosted on the servers of the Department of Physics and Astronomy at the University of Manitoba. 
Public access to the database is granted through a dedicated web site%
\footnote{\url{http://www.physics.umanitoba.ca/snr/SNRcat}}. 
HTML pages are generated on the server by PHP scripts. 
This provides a pre-defined, simple, and almost complete view of the database. 
It~does not yet allow to make custom queries, but it allows for an easy browsing 
of the data without any prerequisite knowledge of the database internals.%
\footnote{Note that the use of some features requires compliance of the user's
browser with current web standards regarding HTML DOM manipulation and CSS styling 
(this can be checked at \url{http://acid3.acidtests.org}).}

\subsection{Main list\label{sub:Main-list}}

The main page is the list of all remnants, with each row corresponding to a single object. 
Clicking anywhere on the row opens the full object record in a new page, with more details and references (see next section).

\subsubsection{Description\label{sub:Description}}

The first columns of the table describe the SNR (identification, environment, main physical properties). 
The first field is the {}``G~number'' that uniquely identifies the object 
(it is made out of its Galactic coordinates~$\left(l,b\right)$, with a fixed 11-character format: $Glll.l\pm bb.b$). 
The other fields from the database are not all shown here individually but in aggregated form. 
The column {}``names'' includes both older identifications and common names, as well as source names as given by high-energy observatories. 
The column {}``context'' contains various information about the SNR environment, 
including associations with compact objects and/or interactions with molecular clouds. 
The columns {}``age'' and {}``distance'' contain all age and distance estimates, respectively%
\footnote{Note that the given age and distance are only our best estimates, and might still be incomplete.}, 
for the shell and/or the pulsar -- noting that the SNR ages are normally determined 
from the SNR dynamics or association with historical events, 
while the pulsar ages are normally estimated from their characteristic age. 
Question marks denote uncertain identifications or associations,
and strikes denote obsolete names or discarded claims 
(all kept in the database for the record). 

The last columns summarize the observational status of the remnants for several modern X-ray and gamma-ray instruments: 
Chandra, XMM-Newton, Suzaku, ROSAT, ASCA, Fermi, AGILE, H.E.S.S., VERITAS, MAGIC, Milagro
(the database also contains records for other observatories, which are all displayed on the individual page of the remnant). 
The current view of each SNR by a given instrument is colour-coded with three cases as follows: 
{}``missed'': the region was observed, but no significant source was detected; 
{}``detected'': a source has been detected but it is not possible to measure its extension; 
{}``extended'': the source is clearly extended, it might show structures or not. 
Note that an empty cell merely means lack of data: 
it might be that the object is not observable with the instrument, 
or that it has not been observed yet, 
or that the results have not been yet disclosed.
In particular, we note that several observatories have made surveys of the Galactic plane (%
ASCA\footnote{\url{http://heasarc.gsfc.nasa.gov/W3Browse/all/ascagps.html}}: \citealt{Sugizaki2001a}, 
H.E.S.S.: \citealt{Aharonian2006a}, 
Milagro: \citealt{Abdo2007b})
or even of the whole sky (%
ROSAT\footnote{\url{http://www.xray.mpe.mpg.de/cgi-bin/rosat/rosat-survey}}: \citealt{Voges1993a}, 
Fermi\footnote{\url{http://fermi.gsfc.nasa.gov/ssc/data/access/}}: \citealt{Fermi-LAT-Collaboration2011a}),
so that unexploited data might still exist on many remnants 
(in the absence of a dedicated publication, they are not indicated in our catalogue).

\subsubsection{Manipulation\label{sub:Manipulation}}

The data can be manipulated to some extent by the user: the main table, although of fixed content, 
provides a number of dynamic features on the client side (if Javascript is enabled). 
The user can re-order the instrumental columns by dragging-and-dropping the header of one column%
\footnote{We use the plugin \texttt{dragtable} available at \url{http://www.danvk.org/wp/dragtable/}}. 
The user can also sort all rows according to an observatory by clicking on the header of its column 
(multiple selections are possible by holding the shift key)%
\footnote{We use the plugin \texttt{Tablesorter} available at \url{http://tablesorter.com/docs/}}. 
Finally, the user can filter the rows in various ways%
\footnote{We use the plugin \texttt{Table Filter} available at \url{http://www.picnet.com.au/picnet_table_filter.html}}: 
by selecting an item in the drop-down list below instruments headers; 
by typing text in the search field below other headers, to filter a given column; 
by typing text in the main search field, to filter the whole table. 
The search in text fields supports logical expressions (AND, OR, NOT), as well as grouping with brackets () and quotes {}``''.
Specific examples are given on the main web page.

\subsection{Individual records\label{sub:Individual-records}}

When the user clicks on the highlighted row on the main page, a new page opens, 
which is the individual record for the selected remnant.
It consists of two tables. 

The first table gives all the properties of the remnant itself: 
identification (unique ID, previous IDs, other names, supernova event), 
location (right ascension and declination of the centroid in J2000, constellation),
environment (general context, plus interactions with clouds), 
age (of the shell, of the pulsar), 
distance (of the shell, of the pulsar),
properties (radio size, radio index, morphology type). 
The table includes references for supernova associations and for molecular cloud associations,
with direct links to ADS records. 

The second table is a list of all known high-energy observations of the remnant. 
Each row corresponds to a single instrument, identified by the energy domain and the observatory name. 
The first field is the source name, as given by the instrument team, if applicable. 
The last fields list all the published papers or official web pages 
presenting relatively new observations with the listed observatory. 
The field {}``resolution'' indicates how well the source is observed: 
{}``missed'', {}``detected'', or {}``extended'' (colour-coded as previously mentioned). 
The field {}``source of emission'' indicates what the observers think is the actual source 
of the high-energy emission observed with the listed observatory. 
Comparing the records of different instruments (which are independent) 
allows to get a better understanding of the underlying object(s). 

The page also provides direct links to Green's catalogue and to the CATS database (when applicable), 
where more details and references can be found for observations at lower energies.

\subsection{Feedback form\label{sub:Feedback-form}}

Feedback can be sent from any page by submitting a form%
\footnote{\url{http://www.physics.umanitoba.ca/snr/SNRcat/SNRform.php}}. 
Users are encouraged to send their comments and suggestions regarding the website,
and to report any incomplete or missing entries.


\section{Statistics\label{sec:Statistics}}
\newcommand{\currentdate}{January~27, 2012}

We provide here simple statistics on the current content of the database.
These will hopefully provide a summary of our current view of Galactic SNRs from a high-energy perspective, 
but note that we cannot claim offering a complete view of the subject. 
First some early or most recent observations may be missing, 
second the census of Galactic SNRs is by no means complete 
and the list is expected to grow with continuing (radio and high-energy) surveys 
and with deep or targeted observations of selected candidates. 
We refer the reader to the work by Green (\citeyear{Green2004b,Green2005a,Green2009b}) 
for SNR population studies and their biases from the radio perspective.

\subsection{Remnants\label{sub:Statistics-Remnants}}

As of \currentdate, the database contains 302 SNRs: 
the 274 objects of Green's catalogue as of March 2009, 
plus 28 objects that were added in the light of recent observations.

\subsubsection{Elements of context}

The astrophysical context of the remnant includes a neutron star (NS) or a NS candidate in 95 cases (31.5\% of the SNRs). 
In 80 cases (84\% of the NSs, 26.5\% of the SNRs) the NS is identified as a pulsar (PSR), 
including 6 anomalous X-ray pulsars (AXPs). 
The 6~AXPs, and 2~soft gamma-ray repeaters (SGRs), 
plus 1~high-magnetic field PSR that displayed a magnetar-like activity, 
imply that 9~magnetar candidates are associated with SNRs (3\% of the SNRs). 
Among the SNRs hosting NSs, 12 (4\% of the SNRs) contain a central compact object (CCO) or CCO candidate, 
an X-ray discovered NS with properties different from the classical rotation-powered pulsars or magnetars 
(3~of them exhibit pulsations, and 1~of them has been also proposed to be an AXP). 
A~relativistic outflow is often associated with the rotation-powered NSs: 
a pulsar wind nebula (PWN) is as a result detected or suggested in 69 cases (23\% of the SNRs). 
Note that these are not a subset of the former population: 
no central object has been reported so far for about 1/3 of the PWNe or PWN candidates
(only 46~SNRs, that is 15\% of the total, are associated with both).
Also note that for many remnants the PWN is actually the only object observed, 
as there is currently no evidence for a shell 
(isolated pulsars with no associated shell or PWN are not listed in this work).
Finally, we warn the reader that not all the objects listed need to be physically associated with the remnant considered
(however, they must be taken into account when interpreting any observations of this region).

Interaction of the shell with a molecular cloud has been reported in 65~cases (21.5\% of the SNRs): 
35 are considered certain (12\%),
11 probable (4\%) and 
19 others possible (6\%)
 --~10~other cases are kept in the database for the record but are considered unlikely.

\subsubsection{Historical events}

The database also contains a table of 14 records of the sighting of a supernova (SN)~\citep{Green2002a,Green2003a}: 
\begin{itemize}

\item 5~certain events (in 1604, 1572, 1181, 1054, 1006);
\item 4~probable events (in 393, 386, 369, 185);
\item 5~spurious events (in 1680, 1592, 1408, 1230, 837).

\end{itemize}
In the SNR table, 14 records refer to one of these supernova events:
\begin{itemize}

\item 4~certain associations (with 4 of the 5 certain SNe): \\
\texttt{G004.5+06.8} with SN~1604 (Kepler), \\
\texttt{G120.1+01.4} with SN~1572 (Tycho), \\
\texttt{G184.6-05.8} with SN~1054 (Crab), \\
\texttt{G327.6+14.6} with SN~1006;

\item 4~probable associations (with the 5th certain SN, and 3 of the 4 probable ones): \\
\texttt{G130.7+03.1} = 3C58 with SN~1181 (the association is still debated),\\
\texttt{G347.3-00.5} = RX~J1713.7-3946 with SN~393, \\
\texttt{G011.2-00.3} with SN~386, \\
\texttt{G315.4-02.3} = RCW~86 with SN~185;

\item 3~suggested associations (with 2 probable SNe), that are no longer favoured: \\
\texttt{G348.5+00.1} = CTA~37A with SN~393, \\
\texttt{G348.7+00.3} = CTA~37B with SN~393, \\
\texttt{G320.4-01.2} = RCW~89 with SN~185;

\item 3~dubious associations (with 3 discarded SNe): \\
\texttt{G111.7-02.1} = Cas~A with SN~1680 (a mistake in Flamsteed's catalogue), \\
\texttt{G069.0+02.7} = CTB~80 with SN~1408 (shown to be a meteor), \\
\texttt{G189.1+03.0} = IC~443 with SN~837 (probably just a nova, and the SNR is much older).

\end{itemize}
Note that, because of the paucity of historical data, the match between the SNR and SN tables is not exact: 
for some SNe no counterpart was proposed (SN~369, SN~1230, SN~1592), 
for others several remnants have been suggested (SN~185, SN~393). 
Also, not all known young SNRs are associated with a historical SN: 
about 32 SNRs in the database (11\%) are estimated to be less than 2000~years old 
(3~times more than convincing SN records over the same period).
\subsection{Observations\label{sub:Statistics-Observations}}

As of \currentdate, the database contains 1149 observational records,
280 radio records (one per SNR, except 22 not present in Green's catalogue
and for which data have to be confirmed), and 869 high-energy records
(including 257 non-detections).

\subsubsection{By observatories}

17 high-energy observatories are present in the database, as shown in Table~\ref{table:records}.

In the X-rays domain, the database aims to be complete for 6~satellites:
ASCA, ROSAT, BeppoSAX, and Chandra, XMM-Newton, Suzaku. 
The first three no longer exist, but they still provide valuable information
(unexploited data from surveys are still regularly published). 
Many remnants (17 according to our census) detected by these instruments
have still not been re-observed with currently operating satellites.
The database also contains observations from 3 other X-ray missions,
not primarily targeted at SNRs, but which can prove useful to pinpoint
a PSR or PWN: INTEGRAL, RXTE, and SWIFT. 

In the $\gamma$-rays domain, the database aims to be complete for 6~observatories: 
AGILE, Fermi, H.E.S.S., VERITAS, MAGIC, and Milagro.
In the MeV range, it also includes a few observations related to nucleosynthesis,
claimed by COMPTEL but not confirmed by INTEGRAL. 
Note that the INTEGRAL observatory may appear separately in \texttt{X} and in \texttt{gamma-MeV} records, 
as these correspond to different instruments, operating at sensibly different energies, and probing completely different physics.
In the GeV range, the observations made with EGRET have not been included,
as associations with SNRs were difficult to establish 
(see \citealt{Torres2003a} for a comprehensive review), 
and data are now superseded by Fermi observations. 
Note that most of the 70 Fermi records are just positional
associations of a detected source with a known remnant, 
the physical connection between the two remains to be proved in most cases. 
Moreover, 15~sources from the bright source list and/or first source catalogue 
have been dismissed in the second source catalogue 
(they are still in the database for the record).
In fact only 11~SNRs have been identified so far by the Fermi collaboration:
W51C, Cas~A, W44, IC443, W28, W49B, W30, Cygnus Loop, RX J1713.7-3946, Tycho, RX J0852.0-4622 = Vela Jr, 
mostly middle-aged SNRs interacting with molecular clouds. 
In addition, 3~PWNe have been identified: the Crab, Vela~X, and MSH~15-52. 
AGILE also provided valuable information for 3~SNRs: IC~443, W28, and W44. 
In the TeV range, observations are dominated by the H.E.S.S. experiment, 
which has detected almost 2/3 of all known sources, 
and has been the only instrument to produce images of SNRs in this energy domain 
(for 5 of them: RX J1713.7-3946, RX J0852.0-4622 = Vela Jr, G353.6-00.7, RCW~86, and SN~1006). 
As for the GeV range, we chose not to include here observations made
by pioneers of the Cherenkov technique, such as the HEGRA experiment.
We include observations made with CANGAROO, although 2~claimed detections
were withdrawn after new observations were made with H.E.S.S.

\subsubsection{By kind of sources}

With Table~\ref{table:sources} we briefly discuss which kind of sources are observed:
do we see the expanding diffuse emission from the remnant, 
or rather the compact object left after the explosion? 
In the first category, we include any kind of emission from the ejecta, 
thermal or (in just a few cases) through nuclear decay, 
as well as any emission from the shock 
(which might involve a nearby molecular cloud in about 20\% of the cases). 
In the second category, we put any kind of identified compact object (PSR, AXP, SGR, NS, CCO), 
as well as any associated jets or wind nebula (PWN). 
In one case (W50) it is actually a micro-quasar,
in two others (RCW 103 and Monoceros Loop) it is a binary candidate. 
It is apparent that the two categories are of comparable size, 
with a slight predominance of the second. 
Sources that do not fall into one of these two broad categories can sometimes be linked to unrelated objects 
such as field stars or dust scattering haloes. 
But many sources (about 20\%) lack any explanation. 
And note that, reversely, there might be competing explanations for a given source of emission, 
or several distinct physical sources of emission (so the four columns are not exclusive).

\begin{table}[H]
\caption{Number of observational records in the database, by energy domain and by instrument 
(numbers are the sum of successful observations and non-detections).}
\label{table:records}

\newcommand{\llwidth}{1cm}
\newcommand{\lwidth}{1cm}
\newcommand{\rwidth}{3cm}
\newcommand{\rrwidth}{3cm}

\begin{tabular}{|c|c|c|>{\centering}m{3cm}|>{\centering}m{3cm}|>{\centering}m{3cm}|}
\hline 
\multicolumn{2}{|c|}{\textbf{domain}} & \textbf{instrument} & \textbf{records by \mbox{instrument}} & \multicolumn{2}{c|}{\textbf{records by domain}}\tabularnewline
\hline
\hline 
\multicolumn{2}{|c|}{\multirow{9}{\lwidth}{X-rays}} & ASCA & $110+{\color{white}00}1=111$ & \multicolumn{2}{c|}{\multirow{9}{\rwidth}{$458+{\color{white}0}17=475$}}\tabularnewline
\cline{3-4} 
\multicolumn{2}{|c|}{} & BeppoSAX & ${\color{white}0}14+{\color{white}00}0={\color{white}0}14$ & \multicolumn{2}{c|}{}\tabularnewline
\cline{3-4} 
\multicolumn{2}{|c|}{} & Chandra & $115+{\color{white}00}0=115$ & \multicolumn{2}{c|}{}\tabularnewline
\cline{3-4} 
\multicolumn{2}{|c|}{} & INTEGRAL & ${\color{white}0}19+{\color{white}00}7={\color{white}0}26$ & \multicolumn{2}{c|}{}\tabularnewline
\cline{3-4} 
\multicolumn{2}{|c|}{} & ROSAT & ${\color{white}0}75+{\color{white}00}0={\color{white}0}75$ & \multicolumn{2}{c|}{}\tabularnewline
\cline{3-4} 
\multicolumn{2}{|c|}{} & RXTE & ${\color{white}0}10+{\color{white}00}4={\color{white}0}14$ & \multicolumn{2}{c|}{}\tabularnewline
\cline{3-4} 
\multicolumn{2}{|c|}{} & Suzaku & ${\color{white}0}31+{\color{white}00}1={\color{white}0}32$ & \multicolumn{2}{c|}{}\tabularnewline
\cline{3-4} 
\multicolumn{2}{|c|}{} & SWIFT & ${\color{white}00}7+{\color{white}00}1={\color{white}00}8$ & \multicolumn{2}{c|}{}\tabularnewline
\cline{3-4} 
\multicolumn{2}{|c|}{} & XMM & ${\color{white}0}77+{\color{white}00}3={\color{white}0}80$ & \multicolumn{2}{c|}{}\tabularnewline
\hline 
\multirow{9}{\llwidth}{$\gamma$-rays} & \multirow{2}{\lwidth}{MeV} & COMPTEL & ${\color{white}00}2+{\color{white}00}0={\color{white}00}2$ & \multirow{2}{\rwidth}{${\color{white}00}2+{\color{white}00}2={\color{white}00}4$} & \multirow{9}{\rwidth}{$154+240=394$}\tabularnewline
\cline{3-4} 
 &  & INTEGRAL & ${\color{white}00}0+{\color{white}00}2={\color{white}00}2$ &  & \tabularnewline
\cline{2-5} 
 & \multirow{2}{\lwidth}{GeV} & AGILE & ${\color{white}00}7+{\color{white}00}0={\color{white}00}7$ & \multirow{2}{\rwidth}{${\color{white}0}77+212=289$} & \tabularnewline
\cline{3-4} 
 &  & Fermi & ${\color{white}0}70+212=282$ &  & \tabularnewline
\cline{2-5} 
 & \multirow{5}{\lwidth}{TeV} & CANGAROO & ${\color{white}00}5+{\color{white}00}6={\color{white}0}11$ & \multirow{5}{\rwidth}{${\color{white}0}75+{\color{white}0}26=101$} & \tabularnewline
\cline{3-4} 
 &  & H.E.S.S. & ${\color{white}0}47+{\color{white}00}1={\color{white}0}48$ &  & \tabularnewline
\cline{3-4} 
 &  & MAGIC & ${\color{white}00}5+{\color{white}0}13={\color{white}0}18$ &  & \tabularnewline
\cline{3-4} 
 &  & Milagro & ${\color{white}00}8+{\color{white}00}0={\color{white}00}8$ &  & \tabularnewline
\cline{3-4} 
 &  & VERITAS & ${\color{white}0}10+{\color{white}00}6={\color{white}0}16$ &  & \tabularnewline
\hline
\multicolumn{2}{|c|}{\textbf{ALL}} & \textbf{TOTAL} & $612+257=869$ & \multicolumn{2}{c|}{$612+257=869$}\tabularnewline
\hline
\end{tabular}

\end{table}

\medskip{}

\begin{table}[H]
\caption{Nature of the high-energy emission source for all observational records in the database 
(for the first three columns, numbers are the sum of confident and uncertain identifications).}
\label{table:sources}

\begin{centering}
\begin{tabular}{|c|c|c|c|c|}
\hline 
 & ejecta / shock & compact object / wind & other (unrelated) & unknown\tabularnewline
\hline
\hline 
X-rays & $184+{\color{white}0}38=222$ & $195+{\color{white}0}54=249$ & ${\color{white}00}8+{\color{white}00}4={\color{white}0}12$ & ${\color{white}0}60$\tabularnewline
\hline 
$\gamma$-rays & ${\color{white}0}24+{\color{white}0}18={\color{white}0}42$ & ${\color{white}0}31+{\color{white}0}15={\color{white}0}46$ & ${\color{white}00}0+{\color{white}00}1={\color{white}00}1$ & ${\color{white}0}60$\tabularnewline
\hline 
TOTAL & $208+{\color{white}0}56=264$ & $226+{\color{white}0}69=295$ & ${\color{white}00}8+{\color{white}00}5={\color{white}0}13$ & $120$\tabularnewline
\hline
\end{tabular}
\par\end{centering}

\end{table}


\section{Perspectives\label{sec:Perspectives}}

We believe that the database will already prove a useful tool as it stands, 
for the interpretation of existing observations as well as for the preparation of future ones, 
for the modelling of individual remnants as well as to do population studies. 
However, it is still work in progress, to be updated regularly following new results 
and any feedback provided to us through the feedback form. 
As new high-energy missions come on-line 
(e.g. NuSTAR, ASTROSAT, eROSITA, and ASTRO-H in the very near future), 
we will update the catalogue with the corresponding new SNR detections or latest studies. 
Finally we discuss here some ideas for future extensions.

\subsection{Wavelength coverage\label{sub:Wavelength-coverage}}

The database was purposely populated with high-energy observations
(with particle acceleration in mind), 
but it was designed in a fairly general way, 
and the long-term goal is to get a full multi-wavelength view of SNRs:
\begin{itemize}

\item Currently, our catalogue relies on radio observations for the identification of SNRs, 
mostly as already compiled by D.~A.~Green. 
It could benefit from an even tighter integration with Green's work, 
by including the details of all radio observations.

\item Energy domains in-between radio and X-rays/$\gamma$-rays 
are also useful for the study of acceleration, 
and these observations shall be included at a later time. 
First, the synchrotron radiation of accelerated electrons might be detected in the optical and IR domains, 
completing the spectrum between the radio continuum and the X-ray cut-off. 
Second, if protons are efficiently accelerated this must impact the dynamics and morphology of the remnant itself, 
which can be probed by observations of the shocked material in the optical and UV domains. 

\end{itemize}

\subsection{Extragalactic coverage\label{sub:Extra-Galactic-coverage}}

Following Green's catalogue, the database was deliberately limited to objects located within our Galaxy. 
It could obviously be extended to extragalactic objects: 
the evolution and detailed morphology of SNRs cannot be studied for the time being in other distant galaxies,
but there are already numerous observations of good quality, 
particularly in the two Magellanic clouds (MC). 
The MC SNRs also have the advantage of not being heavily absorbed (as for the Galactic SNRs) 
and being located at the same known distance.

\subsection{Linking with other databases\label{sub:Linking-with-other-databases}}

For each remnant, our catalogue indicates other related objects, such as molecular clouds or PWNe.
When these objects are catalogued in a dedicated database, 
it becomes possible to offer direct links to the full record of any given object of interest.
In particular, our group is working on building an up-to-date catalogue of PWNe (based on multi-wavelength observations),
and when this new catalogue becomes public we will offer cross-linking between the two databases.


\section*{Acknowledgements}

This research has been supported by Canada Research Chair funds 
from the Natural Sciences and Engineering Research Council of Canada (NSERC), 
the Canada Foundation for Innovation (CFI), 
and the Canadian Institute for Theoretical Astrophysics (CITA). 
This research has also made use of NASA's Astrophysics Data System (ADS) 
and the High-Energy Astrophysics Science Archive Research Center (HEASARC) 
maintained at NASA's Goddard Space Flight Center. \\
We thank Maiko Langelaar (University of Manitoba) for technical support. 


\end{document}